\documentclass[aps,prd]{revtex4}
\usepackage{amsmath}
\usepackage{graphicx}

\begin{document}
\title{Exact Hawking Radiation of Scalars, Fermions, and Bosons Using the Tunneling Method Without Back-Reaction}
\date{February 1, 2011}

\author{Alexandre Yale}
\email{ayale@perimeterinstitute.ca}
\affiliation{Perimeter Institute, 31 Caroline St. N., Waterloo, Ontario N2L 2Y5, Canada}

\begin{abstract}
Hawking radiation is studied for arbitrary scalars, fermions and spin-$1$ bosons, using a tunneling approach, to every order in $\hbar$ but ignoring back-reaction effects. It is shown that the additional quantum terms yield no new contribution to the Hawking temperature.  Indeed, it is found that the limit of small $\hbar$ in the standard quantum WKB approximation is replaced by the near-horizon limit in the gravitational WKB approach.
\end{abstract}
\maketitle

\newcommand{\myeq}[1]{\begin{equation} \begin{split}  #1 \end{split} \end{equation}}

\section{Introduction}
Because it models Hawking radiation as the very physical process of quantum fields tunneling through a barrier, an approach called the tunneling method has recently gained popularity in the field of black hole thermodynamics. This method also provides multiple advantages which go beyond merely providing intuition;  indeed, because it considers Hawking radiation to be a purely local phenomenon, it can be used to study spacetimes with multiple horizons such as embedded black holes in deSitter spacetimes.  Many spacetimes have thus been explored in this way: Kerr-Newman \cite{Jiang2006,Zhang2006}, Black Rings \cite{Zhao2006}, Taub-NUT \cite{Kerner2006}, AdS black holes \cite{Hemming2001}, BTZ \cite{Agheben2005,BTZ,Modak2009}, Vaidya \cite{Vaidya}, dynamical black holes \cite{Cri2007}, Kerr-G\"{o}del \cite{Kerner2007}, deSitter horizons \cite{dS}, constant curvature black holes \cite{Yale2010}, as well as generic weakly isolated horizons \cite{Wu2007}.
\\\\
Moreover, this method is especially powerful in that it allows quantum fields to be considered explicitly.  As such, the Hawking radiation of scalars to every order in $\hbar$ \cite{Majhi1,Majhi4}, spin-$1/2$ fermions to first order \cite{Kerner2008,Kerner2008b,Li2008,Zhang2006,Jiang2006} and later to every order \cite{Majhi2}, higher-spin fermions \cite{Yale2009,Majhi3}, and $U(1)$ gauge bosons to second order \cite{Majhi3} has been investigated.  Expanding on these results, we will consider, to every order, the tunneling of scalars, fermions of any spin, and arbitrary gauge bosons from a generic near-horizon black hole metric.  Some of the results we will present are not technically new; for example, the scalar field was calculated exactly in \cite{Majhi1}.  Nevertheless, we will include them not only for completeness, but also to provide a more thorough derivation and to interpret the results in a slightly different way.
\\\\
The tunneling method comes in two flavours.  The first originates from the works of Volovik \cite{Volovik} and later of Kraus and Wilczek \cite{NullGeo}, who analyzed the radiation process semi-classically by considering modes near the event horizon; the idea was later generalized as a tunneling process by Parikh and Wilczek \cite{PW}.  Near a Schwarzschild horizon, radial null geodesics obey $\frac{dr}{dt} = \dot{r} = \pm (1 - \frac{2M}{r} )$, where the $\pm$ denotes the outgoing and incoming geodesics.  The contribution to the imaginary part of the action comes from two part: a temporal contribution $2E \text{Im} \Delta t = 4 \pi M E$ from the discontinuity of the time coordinate at the horizon \cite{Akhmedov2008b}, and a spacial contribution $\text{Im} \oint p_r dr$, where we integrate along an infinitesimal complex path around the pole at the horizon.  This closed path integral is to be understood as a normalizing process which subtracts the infalling radiation from the outgoing one.  Using the Hamilton equations of motion $p_r =  \int_0^E \frac{dH}{\dot{r}}$, we perform this integral to find $\text{Im} I = \text{Im} E \oint \frac{dr}{\dot{r}} = 4 \pi M E$.  We then relate the tunneling rate to the action by $\Gamma \propto e^{- \text{Im} I} = e^{-8 \pi M E}$, thus finding the correct Hawking temperature of $T = \frac{1}{8 \pi M}$.
\\\\
The second flavour, which we will use throughout this paper, comes from the works of Padmanabhan and his collaborators \cite{PaddyTun}.  This method was initially developed, and later formulated more algorithmically \cite{Agheben2005}, as a means of studying the quantum tunneling of scalar particles through a gravitational barrier.  It is rooted in the WKB approximation and consists of solving the Hamilton-Jacobi equations (thus earning it the nickname of Hamilton-Jacobi method) for a quantum field passing through an event horizon.  For example, for a massless scalar field $\phi = ae^{\frac{-i}{\hbar}I}$, the equations of motion are $\partial_\mu \partial^\mu \phi=0$, of which the leading term is the Klein-Gordon equation $\partial_\mu I \partial^\mu I + {\cal O}(\hbar)= 0$.  This leads to $I = \pm \oint \frac{E dr}{1-2M/r} = 4 \pi i M E$, yielding once more $T = \frac{1}{8 \pi M}$.  A goal of this paper is to show that this $\hbar \rightarrow 0$ approximation, which is also made for fermions and bosons, is unnecessary: since we are only interested in near-horizon physics, we can use $g_{tt} \rightarrow 0$ in place of $\hbar \rightarrow 0$ to retrieve the correct Hawking temperature.  This is somewhat unexpected since taking this limit is so common in semiclassical treatments of Hawking radiation; it means that we can consider this method to be a gravitational analog to the quantum WKB approximation.
\\\\
It is in that sense that our method calculates the temperature exactly to all orders.  We will find that the action for a Klein-Gordon field obeys $\partial_r I= \pm \frac{\partial_t I}{f(r)}$ near the horizon.  This implies two important points: first, since $\partial_t I$ is conserved, we can simply integrate this quantity to find the action without ever having assumed $\hbar$ to be small: as such, our method is exact to every order in $\hbar$.  Second, this formula implies that if we expand the action $I$ in powers of $\hbar$ as $I = \sum \hbar^i I_i$, then each $I_i$ obeys this same equation: $\partial_r I_i= \pm \frac{\partial_t I_i}{f(r)} = \mp \int \frac{E_i}{f(r)}$, where the last equality defined the conserved quantity $E_i = -\partial_t I_i$.  This then means that $\partial_r I = \frac{E_0}{f(r)} \left( 1 + \sum_{i=1}^\infty \frac{\hbar^i E_i}{E_0} \right) = \partial_r I_0 \left( 1 + \sum_{i=1}^\infty \frac{\hbar^i E_i}{E_0} \right)$.
\\\\
Besides $\hbar$, there exists another parameter which is important to the problem of Hawking radiation: the ratio $\frac{E}{M}$ between the energy of the emitted particle and the mass of the black hole; this controls the amount of back-reaction on the black hole. Early in its development, the tunneling method was used to show that this back-reaction modified the thermal nature of the emitted radiation \cite{PW}.  More recently, it was used to study correlations between emitted particles, and may provide a solution to the information puzzle to the lowest order \cite{Zhang2009,Israel2010,Singleton2010}.  This area, however, is beyond the scope of the present paper.  Thus, even though we will expand to every order in $\hbar$, we will ignore back-reaction entirely.
\\\\
Our calculations will be done using a generic near-horizon line element in Schwarzschild-like coordinates:
\myeq{\label{metric} ds^2 = -f(r) dt^2 + \frac{1}{f(r)}dr^2 + d x^2_\bot,}
where $f(r)$ vanishes at the horizon. This form is quite general and does not restrict us to spherically symmetric spacetimes.  For example, in the near-horizon limit at fixed $\theta=\theta_0$, the Kerr metric can be written
\myeq{
ds^2 &= -A(r,\theta) dt^2 + \frac{dr^2}{B(r,\theta)} + C(r,\theta) \left[ d \phi - D(r,\theta)dt \right]^2 + F(r,\theta)d\theta^2 \\
&= -A_r(r_0,\theta_0)(r-r_0) dt^2 + \frac{dr^2}{B_r(r_0,\theta_0)(r-r_0)} + C(r_0,\theta_0) \left[ d \phi - \Omega dt\right]^2,
}
where $\Omega = D(r_0,\theta_0)$ is the angular velocity of the black hole.  This line element is of the form $(\ref{metric})$ up to a redefinition of the $r$ and $\phi$ coordinates.  An unfortunate side-effect of using such a generic metric is a slight misdefinition of the energy, which we will illustrate here for the Kerr spacetime.  From symmetry arguments, we know that the action is of the form $I = -Et + J \phi + W(r,\theta)$.  However, the redefinition of the $\phi$ coordinate $\phi = \chi + \Omega t$ near the horizon means that the action is actually $I = -(E-\Omega J)t + J \chi + W(r,\theta)$.  Therefore, our energy actually corresponds to $(E - \Omega J)$ in standard coordinates.  Moreover, the temperature that we calculate is not redshifted: for an asymptotically flat spacetime, it represents the temperature measured by an observer at infinity.
\\\\
The purpose of this paper is therefore twofold.  First, we will provide a generic treatment of bosons, which have so far only been studied in the Abelian case.  Second, we will show that, ignoring back-reaction, terms of higher order in $\hbar$ do not modify the radiation process; in particular, we find that taking the near-horizon limit has the same effect as sending $\hbar \rightarrow 0$.  In the first three sections, we will analyze, respectively, Klein-Gordon, Rarita-Schwinger and Non-Abelian Yang-Mills fields radiating from the near-horizon metric $(\ref{metric})$.  Then, in Section \ref{sec:temperature}, we calculate the temperature associated with these particles, while, finally, in Section \ref{sec:discussion}, we discuss our conclusions and link our results with similar recent work in the field.

\section{Scalars} \label{sec:scalars}
We begin by considering a massive scalar field $\phi$ which we write $\phi = e^{\frac{-i}{\hbar} I}$.  Although this form is based on the WKB approximation, we will not take the $\hbar \rightarrow 0$ limit which usually accompanies this approximation.  The scalar field $\phi$ obeys the Klein-Gordon equation:
\myeq{ \label{scalar} 0 &= g^{\mu \nu} \left( -\partial_\mu \partial_\nu \phi + \Gamma^\sigma_{\mu \nu} \partial_\sigma \phi \right) + \frac{m^2}{\hbar^2} \phi \\
&= \frac{i}{\hbar}g^{\mu \nu} \left( \partial_\mu \partial_\nu I - \Gamma^\sigma_{\mu \nu} \partial_\sigma I \right) + \frac{1}{\hbar^2} \left( g^{\mu \nu} \partial_\mu I \partial_\nu I + m^2 \right).
}
Upon reaching this stage, one commonly truncates the equation to leading order in $\hbar$ and takes the small-mass limit to retrieve the Hamilton-Jacobi equation $\partial_\mu I \partial^\mu I = 0$.  This truncation, it turns out, is unnecessary, and we can analyze the situation to every order in $\hbar$.  Indeed, we begin by solving the above equation by assuming that $\partial_\mu I$ is real; the imaginary contribution to the action will come from integrating this divergent real quantity around a pole at the event horizon.  We will ultimately find a solution for $I$ which does solve the entire (complex) equation $(\ref{scalar})$, thus justifying this assumption.  Hence, taking the real part of equation $(\ref{scalar})$ and solving for $\partial_r I$, we find
\myeq{  \partial_r I = \pm \sqrt{ \frac{-g^{tt}}{g^{rr}}(\partial_t I)^2 + \frac{-g^{\theta \theta}}{g^{rr}}(\partial_\theta I)^2 + \frac{-g^{\phi \phi}}{g^{rr}}(\partial_\phi I)^2 + \frac{m^2 \hbar^2}{g^{rr}} }. }
Because of the near-horizon symmetry, all of $\partial_t I, \partial_\theta I$ and $\partial_\phi I$ are conserved quantities, and therefore finite.  Moreover, $g^{tt}$ will diverge, such that only the first term inside the square root will contribute.  Thus, we find
\myeq{ \label{scalarFinal} \partial_r I = \pm \sqrt{ \frac{-g^{tt}}{g^{rr}} } \partial_t I .}
As we mentioned earlier, this solution also solves the complex part of equation $(\ref{scalar})$.  Indeed, a simple calculation can show that, for $\partial_r I$ defined by $(\ref{scalarFinal})$, we have
\myeq{ g^{\mu \nu} \partial_\mu \partial_\nu I = g^{rr} \partial_r^2 I = \partial_r I \left( g^{rr} \Gamma^r_{rr} + g^{tt} \Gamma^r_{tt} \right) = g^{\mu \nu} \Gamma_{\mu \nu}^\sigma \partial_\sigma I, }
since most Christoffel symbols vanish near the horizon.
\\\\
This result implies that $I$ will obey the same equation even when we are not taking the limit of $\hbar$ going to zero.  Instead, we take the limit of our metric approaching the event horizon; this is physically justified since that is where the tunneling takes place.  Although more involved mathematically, we will find similar results for fermions and bosons in the following sections.

\section{Fermions} \label{sec:fermions}
It is intuitive that fermions must be emitted at the same temperature as scalar particles.  Indeed, a fermion field of spin $(n+\frac{1}{2})$  is a tensor-valued spinor $\Psi_{\mu_1 \cdots \mu_n a}$ which obeys the Rarita-Schwinger equations.  Although these are commonly written, for a spin-$3/2$ field, as
\myeq{\left( \epsilon^{\sigma \nu \rho \mu} \gamma^5 \gamma_\nu \partial_\rho - i m \sigma^{\sigma \mu} \right) \Psi_{\mu a},  }
it is, for our purposes, much more enlightening to write them as
\myeq{\left( -i \gamma^\mu D_\mu + m \right) \Psi_{\mu_1 \cdots \mu_n a} &= 0 \\
\gamma^{\mu_1} \Psi_{\mu_1 \cdots \mu_n a} &= 0,
}
where
\myeq{D_\mu = \partial_\mu - \frac{1}{8} \Gamma^\alpha_{\mu \nu}g^{\nu \beta}[\gamma_\alpha,\gamma_\beta];}
this is the form in which they were originally studied \cite{Rarita1941}.  In flat space, we can multiply by $(i \gamma^\nu D_\nu + m)$ to notice that each component of $\Psi_{\mu_1 \cdots \mu_n a}$ must obey the Klein-Gordon equation. Additional constraints then relate the components of $\Psi_{\mu_1 \cdots \mu_n a}$ with one another: the Dirac equation relates the Dirac indices $a$, while the other Rarita-Schwinger equations relate the higher-spin Lorentz indices $\mu_1 \cdots \mu_n$.  Since we write the field as $\Psi_{\mu_1 \cdots \mu_n a} = a_{\mu_1 \cdots \mu_n a} e^{\frac{i}{\hbar} I}$ and are interested in the action, these extra relations play no role in the calculation of the Hawking temperature, and therefore fermions must be emitted at the same temperature as scalar particles.
\\\\
The Hawking temperature of fermions has already been calculated to leading order in $\hbar$ \cite{Kerner2008,Li2008,Zhang2006,Jiang2006,Yale2009} and, more recently, to every order \cite{Majhi2,Majhi3} for the massless case.  We will perform the calculation for the massive arbitrary-spin case to every order in $\hbar$, and will retrieve the results from \cite{Majhi3}, albeit with an additional term which will not contribute to the Hawking temperature.  The appearance of this term is due to our calculations being more thorough than previous ones, as we attempt to fill the gaps left behind by previous works. Moreover, our work has a slightly different interpretation than that of \cite{Majhi3}, as we will discuss in Section \ref{sec:discussion}.
\\\\
The Dirac equation implies that
\myeq{ \label{dirac} 0 = -i\gamma^\mu \left( \partial_\mu a_{\mu_1 \cdots \mu_n a} - \frac{i}{\hbar}a_{\mu_1 \cdots \mu_n a}\partial_\mu I \right) + ma_{\mu_1 \cdots \mu_n a} + \frac{i}{8} \gamma^\mu g^{\nu \beta}\Gamma^\alpha_{\mu \nu}[\gamma_\alpha,\gamma_\beta]a_{\mu_1 \cdots \mu_n a},}
while the other Rarita-Schwinger equations will simply relate the various $\mu_i$ indices and will have no effect on the action; more details can be found in \cite{Yale2009}.  We define the vierbein $e^I_\mu$ so that $e^I_\mu e^J_\nu \eta^{\mu \nu} = g^{IJ}$; for the metric $(\ref{metric})$, this means $e_a^b = \sqrt{|g^{aa}|} \delta_a^b$.  We also define the Dirac matrices $\gamma^I = e^I_\mu \hat{\gamma}^\mu$, where the $\hat{\gamma}^\mu$ represent the flat-space $\gamma$ matrices in Majorana representation:
\myeq{ \gamma^0=  \left( \begin{array}{cc} 0 & 1 \\ -1 & 0 \end{array} \right) \hspace{1.cm}
\gamma^i=  \left( \begin{array}{cc} 0 & \sigma^i \\ \sigma^i & 0 \end{array} \right),
}
where the $\sigma^i$ are the standard Pauli matrices.  In particular, we have $\left\{ \gamma^I,\gamma^J \right\} = 2 e^I_\mu e^J_\nu \eta^{\mu \nu} = 2g^{IJ}$.  Near the horizon, the metric only depends on the radial coordinate, which means that $\partial_\mu$ for $\mu \neq r$ represents a Killing vector.  This then implies that $a_{\mu_1 \cdots \mu_n a}$ can only be a function of $r$: the dependence of the fermion field on the other coordinates is restricted to a phase (such as $e^{iEt}$).  Therefore, $\gamma^\mu \partial_\mu a_{\mu_1 \cdots \mu_n a} = \gamma^r \partial_r a_{\mu_1 \cdots \mu_n a} = \sqrt{g^{rr}} \hat{\gamma}^r \partial_r a_{\mu_1 \cdots \mu_n a} = 0$, since $g^{rr}$ vanishes while $a_{\mu_1 \cdots \mu_n a}$ remains finite.  Hence, near the horizon, the Dirac equations become a system of equations linear in $a_{\mu_1 \cdots \mu_n a}$:
\myeq{ 0 = \left( \frac{-1}{\hbar}\gamma^\mu \partial_\mu I + m + \frac{i}{8} \gamma^\mu g^{\nu \beta}\Gamma^\alpha_{\mu \nu}[\gamma_\alpha,\gamma_\beta]\right) a_{\mu_1 \cdots \mu_n a} .}
Reading this as a matrix equation in the Dirac indices, it is obvious that $\left( \frac{-1}{\hbar}\gamma^\mu \partial_\mu I + m + \frac{i}{8} \gamma^\mu g^{\nu \beta}\Gamma^\alpha_{\mu \nu}[\gamma_\alpha,\gamma_\beta]\right)$ being invertible would imply $a_{\mu_1 \cdots \mu_n a} = 0$.  Thus, we demand
\myeq{ 0 &= \text{Det} \left( \frac{-1}{\hbar}\gamma^\mu \partial_\mu I + m + \frac{i}{8} \gamma^\mu g^{\nu \beta}\Gamma^\alpha_{\mu \nu}[\gamma_\alpha,\gamma_\beta]\right) \\
&= \text{Det} \left( \begin{array}{c c c c}
-\hbar m & 0 & A & B \\
0 & -\hbar m & C & D \\
-D & B & -\hbar m & 0 \\
C & -A & 0 & -\hbar m
\end{array} \right) \\
&=  (AD-BC)^2 + 2m^2(AD-BC) + m^4
}
where we defined
\myeq{
A &= \sqrt{-g^{tt}}\partial_t I + \sqrt{g^{rr}} \partial_r I + \frac{i \hbar}{4} \sqrt{g^{tt}}(g^{tt} g^{rr})^{3/2} g_{tt,r} \\
B &= \sqrt{g^{\theta \theta}} \partial_\theta I + i \sqrt{g^{\phi \phi}} \partial_\phi I \\
C &= \sqrt{g^{\theta \theta}} \partial_\theta I - i \sqrt{g^{\phi \phi}} \partial_\phi I \\
D &= \sqrt{-g^{tt}}\partial_t I - \sqrt{g^{rr}} \partial_r I - \frac{i \hbar}{4} \sqrt{g^{tt}}(g^{tt} g^{rr})^{3/2} g_{tt,r}.
}
As we approach the horizon, $A$ and $D$ diverge, such that the last terms do not contribute.  We ultimately find $AD=0$, which implies
\myeq{ \label{fermionFinal} \partial_r I = \sqrt{ \frac{ -g^{tt} }{ g^{rr} } } \left(\pm  E - \frac{i \hbar}{4}(g^{tt}g^{rr})^{3/2}g_{tt,r} \right) .}
\\\\
As discussed in \cite{Kerner2008}, studying fermions provides us with insight which is absent from the scalar case: a direct meaning for the $\pm$ in $(\ref{fermionFinal})$.  Indeed, consider the massless spin-$1/2$ case, where the fermion field is $\Psi_a = a_ae^{\frac{i}{\hbar}I}$.  The spin-up case corresponds to
\myeq{ a_a = \left[ \begin{array}{c} \xi_+ \alpha \\ \xi_+ \beta \end{array} \right]
= \left[ \begin{array}{c} \alpha \\ 0 \\ \beta \\ 0 \end{array} \right] ,}
where $\xi_+$ is the positive-spin eigenvector of $\sigma_r$.  Then, combining equations $(\ref{dirac})$ and $(\ref{fermionFinal})$, we find that either $A=0$ or $B=0$.  If $A=0$, $a_a$ will be an eigenvector of $\gamma^5$ with positive eigenvalue and therefore right-handed, whereas if $B=0$, $a_a$ will be left-handed.  Thus, since they have the same spin, the two solutions of $(\ref{fermionFinal})$ correspond to particles of opposite momenta: one is falling into the black hole whereas the other is outgoing.

\section{Bosons} \label{sec:bosons}
Although very little attention has been given to bosons using the tunneling method, the emission of a $U(1)$ field from a generic black hole has recently been considered up to second order in $\hbar$ \cite{Majhi3}.  We will here expand on this result to find an exact formula for $\partial_r I(r)$ for an arbitrary non-Abelian Yang-Mills theory.  We therefore consider a vector field $A_\mu^a = a_\mu^a e^{\frac{-i}{\hbar} I}$ obeying the equations of motion
\myeq{ 0 &= \nabla^\nu F_{\mu \nu} \\
&= g^{\nu \alpha} \left[ \partial_\alpha F_{\mu \nu}^a - \Gamma_{\alpha \mu}^\lambda F_{\lambda \nu}^a - \Gamma_{\alpha \nu}^\lambda F_{\mu \lambda}^a + gf^{abc} A_\nu^b F_{\alpha \mu}^c \right] ,
}
where we defined $F_{\mu \nu}^a = \partial_\mu A_\nu^a - \partial_\nu A_\mu^a + gf^{abc}A_\mu^b A_\nu^c$.  Expanding this according to $A \propto a e^{\frac{i}{\hbar}I}$, we find the expression
\myeq{ \label{long1}
0 = g^{\nu \alpha} \bigg[
& \partial_\alpha \partial_\mu a_\nu^a - \frac{i}{\hbar}\partial_\alpha a_\nu^a \partial_\mu I - \frac{1}{\hbar^2}a_\nu^a \partial_\alpha I \partial_\mu I - \frac{i}{\hbar}a_\nu^a \partial_\alpha \partial_\mu I - \frac{i}{\hbar}\partial_\mu a_\nu^a \partial_\alpha I \\
&-\partial_\alpha \partial_\nu a_\mu^a + \frac{i}{\hbar}\partial_\alpha a_\mu^a \partial_\nu I + \frac{1}{\hbar^2}a_\mu^a \partial_\alpha I \partial_\nu I + \frac{i}{\hbar} a_\mu^a \partial_\alpha \partial_\nu I + \frac{i}{\hbar}\partial_\nu a_\mu^a \partial_\alpha I \\
&+ gf^{abc}A_\nu^c \left( \partial_\alpha a_\mu^b - \frac{i}{\hbar}a_\mu^b \partial_\alpha I \right) \\
&+ gf^{abc}A_\mu^b \left( \partial_\alpha a_\nu^c - \frac{i}{\hbar}a_\nu^c \partial_\alpha I \right) \\
&-\Gamma^\lambda_{\alpha \mu}\left( \partial_\lambda a_\nu^a - \frac{i}{\hbar}a_\nu^a\partial_\lambda I - \partial_\nu a_\lambda^a + \frac{i}{\hbar}a_\lambda^a \partial_\nu I + gf^{abc}A_\lambda^b A_\nu^c \right) \\
&+\Gamma^\lambda_{\alpha \nu}\left( \partial_\lambda a_\mu^a - \frac{i}{\hbar}a_\mu^a \partial_\lambda I - \partial_\mu a_\lambda^a + \frac{i}{\hbar}a_\lambda^a \partial_\mu I + gf^{abc}A_\lambda^b A_\mu^c \right) \\
&+ gf^{abc}A_\nu^b \left( \partial_\alpha a_\mu^c - \frac{i}{\hbar}a_\mu^c \partial_\alpha I - \partial_\mu a_\alpha^c + \frac{i}{\hbar}a_\alpha^c \partial_\mu I + gf^{cde}A_\mu^d A_\alpha^e \right) \bigg].
}
We first simplify this expression by fixing the gauge:
\myeq{ 0 &= \nabla_\mu A^{a \mu} \\
&= g^{\nu \alpha} \left[ \partial_\alpha a_\nu^a - \frac{i}{\hbar}a_\nu^a \partial_\alpha I - \Gamma^\lambda_{\nu \alpha}a_\lambda^a \right];
}
then, $(\ref{long1})$ becomes
\myeq{ \label{long2}
0 = g^{\nu \alpha} \bigg[
& \partial_\alpha \partial_\mu a_\nu^a  - \frac{i}{\hbar}a_\nu^a \partial_\alpha \partial_\mu I - \frac{i}{\hbar}\partial_\mu a_\nu^a \partial_\alpha I \\
&-\partial_\alpha \partial_\nu a_\mu^a + \frac{i}{\hbar}\partial_\alpha a_\mu^a \partial_\nu I + \frac{1}{\hbar^2}a_\mu^a \partial_\alpha I \partial_\nu I + \frac{i}{\hbar} a_\mu^a \partial_\alpha \partial_\nu I + \frac{i}{\hbar}\partial_\nu a_\mu^a \partial_\alpha I \\
&+ gf^{abc}A_\nu^c \left( \partial_\alpha a_\mu^b - \frac{i}{\hbar}a_\mu^b \partial_\alpha I \right) \\
&+ gf^{abc}A_\mu^b \left( \partial_\alpha a_\nu^c - \frac{i}{\hbar}a_\nu^c \partial_\alpha I \right) \\
&-\Gamma^\lambda_{\alpha \mu}\left( \partial_\lambda a_\nu^a - \frac{i}{\hbar}a_\nu^a\partial_\lambda I - \partial_\nu a_\lambda^a + \frac{i}{\hbar}a_\lambda^a \partial_\nu I + gf^{abc}A_\lambda^b A_\nu^c \right) \\
&+\Gamma^\lambda_{\alpha \nu}\left( \partial_\lambda a_\mu^a - \frac{i}{\hbar}a_\mu^a \partial_\lambda I - \partial_\mu a_\lambda^a + gf^{abc}A_\lambda^b A_\mu^c \right) \\
&+ gf^{abc}A_\nu^b \left( \partial_\alpha a_\mu^c - \frac{i}{\hbar}a_\mu^c \partial_\alpha I - \partial_\mu a_\alpha^c + \frac{i}{\hbar}a_\alpha^c \partial_\mu I + gf^{cde}A_\mu^d A_\alpha^e \right) \bigg].
}
We now focus on the $\mu=t$ equation.  Setting the time derivatives of $a_\mu$ to zero and simplifying slightly, we find
\myeq{
0 &= g^{tt} \left[ \frac{1}{\hbar^2}a_t^a (\partial_t I)^2 + \frac{i}{2\hbar} g^{rr}g_{tt,r} a_r^a \partial_t I \right] \\
&+ g^{rr} \bigg[
- \frac{i}{\hbar}a_r^a \partial_r \partial_t I
-\partial_r \partial_r a_t^a + \frac{i}{\hbar}\partial_r a_t^a \partial_r I + \frac{1}{\hbar^2}a_t^a \partial_r I \partial_r I + \frac{i}{\hbar} a_t^a \partial_r \partial_r I + \frac{i}{\hbar}\partial_r a_t^a \partial_r I \\
&+ gf^{abc}A_r^c \partial_r a_t^b
+ gf^{abc}A_t^b  \partial_r a_r^c\\
&-\Gamma^t_{rt}\left( - \frac{i}{\hbar}a_r^a\partial_t I - \partial_r a_t^a + \frac{i}{\hbar}a_t^a \partial_r I + gf^{abc}A_t^b A_r^c \right) \\
&+\Gamma^r_{rr}\left( \partial_r a_t^a - \frac{i}{\hbar}a_t^a \partial_r I + \frac{i}{\hbar} a_r^a \partial_t I + gf^{abc}A_r^b A_t^c \right) \\
&+ gf^{abc}A_r^b \left( \partial_r a_t^c - \frac{i}{\hbar}a_t^c \partial_r I + \frac{i}{\hbar}a_r^c \partial_t I + gf^{cde}A_t^d A_r^e \right) \bigg] \\
&+ (\cdots),
}
where the $(\cdots)$ refers to the $\theta$ and $\phi$ sectors of the equation.  We omitted those terms since they will not contribute to $\partial_r I$ near the horizon, because they will remain finite while other terms, such as $g^{tt} (\partial_t I)^2$, diverge.  Using the fact that the $a_\mu^a$ are normalized to be finite everywhere near the horizon (such that, for example, $g^{rr} a_\mu \rightarrow 0$), and looking only at the real part of this equation \footnote{We are solving for $\text{Im} I$ by integrating $\partial_r I$ around a pole at the horizon.  Thus, only real divergent terms can contribute.}, we find
\myeq{
0 &= g^{tt} \left[ \frac{1}{\hbar^2}a_t^a (\partial_t I)^2 \right] \\
&+ g^{rr} \left[
\frac{1}{\hbar^2}a_t^a \partial_r I \partial_r I + \Gamma_{rt}^t \partial_r a_t^a - \Gamma^t_{rt} gf^{abc}A_t^b A_r^c - g\frac{i}{\hbar}f^{abc}A_r^b a_t^c \partial_r I \right].
}
Since $\Gamma^t_{rt}=\frac{1}{2}g^{tt}g_{tt,r}$, then $g^{tt} \gg g^{rr} \Gamma^t_{rt}$ such that the two middle terms in the square brackets vanish.  Our expression is therefore a quadratic polynomial in $\partial_r I$, which we will denote $0=ax^2 + bx + c$ where $a, b$ and $\frac{1}{c}$ all go to zero at the same speed.  Then, since $\frac{b}{a}$ is finite while $\frac{c}{a}$ diverges, we find $x = \pm \sqrt{ \frac{-c}{a}}$.  Hence:
\myeq{ \partial_r I = \pm \sqrt{ \frac{-g^{tt}}{g^{rr}}} \partial_t I .}

\section{Temperature} \label{sec:temperature}
Calculating the Hawking temperature from $\partial_r I$ has long been a contested issue, as many questions surrounding the covariance of the method have been raised. In particular, it appeared to yield different Hawking temperatures, differing by a factor of two \cite{Pilling2008}, depending on the coordinate system used.  Now that the tunneling method has matured and become better understood, it is generally felt that we have a good handle on this issue.
\\\\
While many potential techniques have been proposed \cite{Mitra2006,Akhmedov2008,Stotyn2009,Majhi5,Akhmedov2006,Akhmedov2006b,Chowdhury2006}, we will here calculate the temperature using the approach recently summarized in \cite{Gill2010} which, in particular, assumes that the tunneling rate follows a thermal distribution.  The temperature gets two contributions: one from the integration over the radial coordinate and one from the discontinuity in the time coordinate:
\myeq{ \label{temp} T_H = \frac{E}{\text{Im} \left( \int \partial_r I_+ - \int \partial_r I_- + 2E \Delta t \right)}.}
We begin by calculating the contribution from $\partial_r I$.  In all cases, defining the energy as $E = -\partial_t I$, we have $\partial_r I = \frac{1}{f(r)} \left(\pm  E + C \right)$ for some finite function $C$.  It is clear, then, that
\myeq{ \int \partial_r I_+ - \int \partial_r I_- &= \int \frac{1}{f(r)} \left( E + C \right) - \int \frac{1}{f(r)} \left( -E + C \right) \\
&= E \oint \frac{1}{f(r)}  \\
&= \frac{E}{f'(r_H)} 2 \pi i
}
Next, we find the contribution from the discontinuity in the time coordinate, $\Delta t$, across the horizon.  The metric $(\ref{metric})$ corresponds to an accelerated observer in flat space who follows the path
\myeq{ t_{\text{out}} &= \frac{ \sqrt{f(r)}}{a} \sinh(a t) \hspace{2.cm} t_{\text{in}} = \frac{ \sqrt{-f(r)}}{a} \cosh(a t) \\
x_{\text{out}} &= \frac{ \sqrt{f(r)}}{a} \cosh(a t) \hspace{2.cm} x_{\text{in}} = \frac{ \sqrt{-f(r)}}{a} \sinh(a t),
}
where the ``in'' and ``out'' subscripts refer to whether we are considering $r \leq r_H$ or $r > r_H$, and where $a = \frac{f'(r)}{2}$.  Thus, as the horizon is crossed, we need $t \rightarrow t - \frac{i \pi}{2a}$, so $\Delta t = \frac{i \pi}{f'(r)}$.  Hence, $2E \Delta t = \frac{2E i \pi}{f'(r)}$ and, from $(\ref{temp})$, we get the Hawking temperature
\myeq{ T_H = \frac{f'(r)}{4\pi} }
for every type of particle.  This agrees with the temperature commonly found in the literature, which is usually calculated only to leading order in $\hbar$.

\section{Discussion} \label{sec:discussion}
We've completed the study of spin-$1$ bosons, initiated in \cite{Majhi3}, by extending it to the non-Abelian case, by giving proper physical motivation for a number of terms dropping out, and by calculating all higher-order terms.  Combined with previous results for scalars and fermions, this finally confirms that Hawking radiation is independent of the type of particle involved.  Although our results show that bosons are emitted at the Hawking temperature regardless of the symmetries of the underlying theory, we needed to fix the gauge ($\nabla_\mu A^\mu = 0$) in order to perform the calculations.
\\\\
We've also shown that the $\hbar \rightarrow 0$ limit is unnecessary when using the tunneling method.  Indeed, since the method is highly local in considering the emission of a field from a pole at the horizon, we are forced to take the limit $r \rightarrow r_H$.  It is therefore natural to take $g_{tt} \rightarrow 0$ instead of $\hbar \rightarrow 0$, such that the tunneling method can truly be understood as the gravitational analog to the quantum WKB method.  There are, however, some important drawbacks to this approach.  First, by assuming that there is no back-reaction, we are drastically restricting quantum processes which may affect the radiation process.  Second, one might expect to see grey-body corrections; the fact that these are missed by the tunneling approach makes this method suspect \footnote{The author would like to thank T Padmanabhan for bringing this point to his attention.}.
\\\\
It is also important to distinguish our results from those of Majhi et al.\ \cite{Majhi1,Majhi2,Majhi3}, who have recently calculated non-zero contributions from the Hawking temperature coming from higher-order terms in the tunneling method.  This mismatch is simply a consequence of differing definitions of energy, which we can illustrate using the free scalar field.  For this case, our calculations yield equation $(\ref{scalarFinal})$, which is also found in \cite{Majhi1}.  We define the energy as $E = -\partial_t I$, and therefore find no additional contributions to the Hawking temperature.   On the other hand, \cite{Majhi1} defines the energy as $E = -\partial_t I_0$ where $I_0$ is the leading term in the action $I = \sum \hbar^i I_i$; this then produces higher-order corrections.  We are not the first to question these corrections; indeed, \cite{Wang2010,Mitra} have pointed out that they are caused by an odd definition of the field's energy and concluded that the Hawking temperature is not modified by higher-order terms.  Moreover, in previous works on this topic, terms with no explicit dependence on the action (such as the last term in our equation $(\ref{fermionFinal})$) are automatically set to zero; we've filled this gap by providing the necessary justification as to how each term cannot contribute to the Hawking temperature.
\\\\
In conclusion, we have calculated the temperature associated with the emission rate of every type of scalars, fermions, and spin-$1$ bosons from a generic black hole spacetime and to every order in $\hbar$ in the tunneling approach, as long as there is no back-reaction.

\acknowledgments{
This work was supported by the Natural Sciences and Engineering Research Council of Canada.  The author would like to thank Ross Diener, Nima Doroud, as well as the reviewer for insightful comments on the manuscript.
}

\end{document}